# Enhanced and controllable reflected group delay based on Tamm surface plasmons with Dirac semimetals

QiwenZheng[1], WenguangLu[2], ShenpingWang[1], XinminZhao[1,*], and Leyong Jiang [1,†]

[1]*School of Physics and Electronics, Hunan Normal University, Changsha 410081, China;*

[2] *School of Electronic Science and Engineering, National University of Defense Technology Changsha 410073, China.*

Corresponding Author: *zhaoxinmin@hunnu.edu.cn and †jiangly28@hunnu.edu.cn

## Abstract

In this paper, the reflected group delay from a multilayer structure where Dirac semimetal is coated on one-dimensional photonic crystal (1D PC) separated by a spacer layer is investigated theoretically. It is shown that the group delay of reflected beam in this structure can be significant enhanced negatively and can be switched from negative to positive. The enhanced group delay originates from the steep phase change caused by the excitation of Tamm plasmons at the interface between the Dirac semimetal and spacer layer. It is clear that the positive and negative group delay can be actively tuned through the Fermi energy and the relaxation time of the Dirac semimetal. We believe this enhanced and tunable delay scheme is promising for fabricating optical delay devices and other applications at middle infrared band.

# 1. Introduction

Group delay generally indicates the speed of the phase change relative to angular frequency when the electromagnetic wave pulse passes through a medium or structure ,it is a typical optical phenomenon in the transmission process of optical pulse [1,2]. Since the delay time of reflected and transmitted pulses can be manipulated from subluminal to superluminal by controlling the dispersive properties of the medium [3], people realize that the research on group delay plays an important role in many fields, such as optical communications [4,5], delay lines [6],efficient optical storage [7]and all-optical switching [8]. Therefore, researchers are committed to the ways of group delay enhancement and regulation in various structures. For example, the delay phenomenon in Otto structure [9], Bragg mirror structure [10] and Fabry- Perot cavity structure [11] have attracted extensive attention. In addition, Wang et al. studied the phenomenon of reflected optical pulse group delay of weakly absorbing dielectric slab, and they obtained negative group delay of -4 ps [12]; Yao et al. studied the tunable group delay in Fabry- Perot cavity, and regulated the negative group delay by adjusting the polarization azimuthal angle of the incident pulse [13]. Although people have studied many delay phenomena based on various structures and proposed many methods to enhance group delay, it is still a challenging work to study the delay approaches and schemes with large group delay. The exploration of micro-nano structures with large group delay combined with new materials has attracted people's attention. Taking the representative graphene in two-dimensional materials as an example, it has obvious advantages in realizing enhanced delay phenomenon due to its excellent photoelectric characteristics [14, 15]. Researchers have studied many ways to enhance and regulate the reflected group delay of optical pulse in graphene based micro- nano structures. For example, Li et al. improved Otto structure with graphene embedded, obtained the reflected group delay by exciting surface plasmons (SPPs) and parameter optimization. They can realize the flexible conversion between positive and negative group delays at the same time [9]; Wang et al. studied the regulation characteristics of the reflected group delay in both mechanisms of resonances and the excitations of surface plasmon resonances in graphene based layered systems, by adjusting the Fermi level and temperature of graphene, the group delay under both mechanisms can realize positive and negative conversion [16]; Xu et al. studied the tunable bistable reflected group delay based on graphene nonlinear surface plasmon resonance, and verified that the bistable reflected group delay can be adjusted by properly adjusting the Fermi level and electron relaxation time of graphene [17]. It is not difficult to see that the group delay of micro- nano structure's enhancement and regulation approach based on two-dimensional materials will be one of the feasible directions to realize practical delay devices. It is still the direction of people's efforts to study the optical delay approaches with large group delay, easier adjustment and simple structure.

In terms of mechanism, optical Tamm states (OTS), as a lossless interface mode localized at the boundary of two different periodic dielectric structures, also has advantages in the field of enhanced group delay [18]. Compared with surface plasmons, OTS does not need a specific incident angle to achieve wave vector matching, it can be excited even at vertical incidence [19]. At present, the structure which excited OTS mainly depends on one-dimensional crystal hetero structure [20] and metal / graphene Bragg mirror structure (M-DBR) [21]. In addition, when OTS is excited in the photonic band gap (PBG), it will be accompanied by a strong local field enhancement effect, and OTS corresponds to the reflection resonance peak in the PBG in the reflection spectrum [22]. These characteristics of OTS make it widely studied in absorption characteristics [23], nonlinear Kerr effect [24] and optical sensor [25]. At the same time, due to its simple excitation structure and the excitation is accompanied by drastic change of phase, it provides an effective way to design optical devices with large group delay. In recent years, a new quantum matter – bulk Dirac semimetals (BDS ) [26] has attracted people's attention. This material can be roughly regarded as "3D graphene", because it is similar to grapheme in many ways, its Fermi level can be adjusted by chemical doping and changing the bias voltage [27], so as to change its dielectric constant [28, 29]. At present, BDS has been applied in the fields of light wave absorbers and optical sensors. For example, Liu et al. proposed a terahertz tunable narrowband absorber based on BDS, which realizes the dynamic regulation of the absorption frequency in terahertz band by adjusting the Fermi level [26]; Luo et al. proposed a high sensitivity tunable terahertz sensor based on BDS to solve the problem of low figure of merit (FOM), it is verified that the FOM of the sensor reaches 813 [30]. At the same time, BDS shows metal response at frequencies lower than Fermi level and dielectric response at frequencies higher than Fermi level [31]. Therefore, adding it to a specific optical structure to excite OTS is a way to obtain adjustable large group delay. In order to control the reflected group delay, a composite structure composed of BDS and Bragg mirror is proposed in this paper. The OTS is excited by BDS and one-dimensional photonic crystal (BDS- 1D PC) structure and combined with the optimization of structure's parameters to obtain a large negative group delay. We believe that the tunable large group delay based on BDS composite structure can find potential applications in photoelectric detection, sensing, communication and other fields.

## 2. Theoretical Model and Method

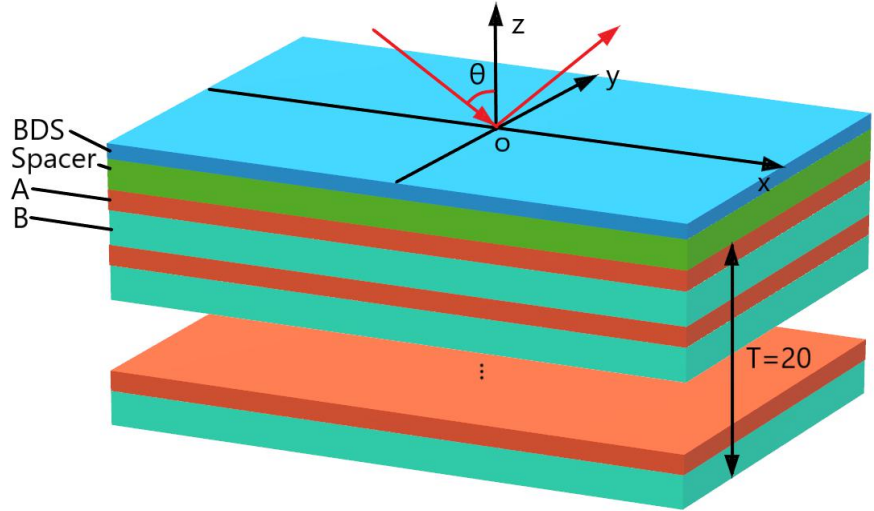

Fig.1. Schematic diagram of one-dimensional photonic crystal structure containing

bulk Dirac semimetals

We consider a composite structure composed of BDS and 1D PC to excite the OTS. The spacer layer is placed between BDS and photonic crystal, as shown in Figure 1. 1D PC is composed of alternating T cycles of dielectric layer A and dielectric layer B. We set the refractive index of dielectric layer A with thickness $d_A$ as 1.47 and that of dielectric layer B with thickness $d_B$ as 1.9. Their refractive indices in terahertz band can be obtained from poly (4-methyl-1-pentene) (TPX) and $SiO_2$, respectively. In the following calculation, we set the period T = 20, the central wavelength of the photonic crystal reflection band is $\lambda_C = 60$ μm, and the thickness of media A and B meets $d_A = \lambda_C / 4n_A$, $d_B = \lambda_C / 4n_B$. In order to simplify the calculation, only the case of TE polarization is considered.

In the above BDS-1D PC composite structure, BDS is the key to dynamically adjust the excitation wavelength of OTS. Under the limitation of long wavelength, the photoelectric characteristics of BDS can be expressed by conductivity $\sigma$. When low

temperature conditions T $E_F$ are met, the conductivity of BDS can be approximately expressed as [31]:

$$\mathrm{Re}(\sigma(f)) = \frac{e^2}{\hbar} \frac{gk_F}{24\pi} \Omega(f)\theta(\Omega(f)-2) \qquad (1),$$

$$\mathrm{Im}(\sigma(f)) = \frac{e^2}{\hbar} \frac{gk_F}{24\pi} \left[ \frac{4}{\Omega(f)} - \Omega(f) ln\left(\frac{4\varepsilon_c^2}{|\Omega^2(f)-4|}\right) \right] \qquad (2),$$

where $g$ is the degeneracy factor. In this paper, we set $g = 40$, the Fermi velocity $v_F$ as $10^6$ m/s, the Fermi level is $E_F$ and the Fermi momentum is $k_F = E_F/\hbar v_F$, $\hbar$ as the reduced Planck constant, $\Omega(f) = 2\pi\hbar f/E_F + i\hbar\tau^{-1}/E_F$. Where $\tau = \mu E_F/v_F^2$ is the electron relaxation time, $\mu$ is the carrier mobility, $\varepsilon_c = E_c/E_F$, where $E_c$ is the cut-off energy, once this energy is exceeded, the Dirac spectrum will no longer be linear [26]. Based on the above formula, the relationship between the dielectric constant of BDS and its conductivity can be expressed as:

$$\varepsilon_{BDS} = \frac{i*(Re(\sigma(f))+i*Im(\sigma(f)))}{2\pi\varepsilon_0 f}, \qquad (3)$$

where $\varepsilon_0$ is the absolute dielectric constant. It can be seen from the above expression that the Fermi level plays a central role in characterizing the conductivity of BDS, and the Fermi level $E_F$ can be controlled by applying external voltage, which provides an effective way to regulate the reflected group delay of the composite structure which surface covered with BDS.

In order to calculate the reflected group delay of the above structure, we need to obtain the reflection characteristics of the incident wave on the surface of the structure. We use the transfer matrix method to calculate the reflection coefficient of BDS-1D

PC structure. The interaction between medium and light wave can be completely determined by the characteristic matrix of its medium layer. Based on the boundary conditions, the characteristic matrix $M_j$ of single-layer medium can be obtained:

$$M_j = \begin{bmatrix} \cos\delta_j & \dfrac{i}{\eta_j}\sin\delta_j \\ i\eta_j \sin\delta_j & \cos\delta_j \end{bmatrix}, \quad (4)$$

where $\eta_j = K_j / \varepsilon_j$, for the structure of multi-layer medium, the characteristic matrix of multi-layer medium $M_{total}$ can be obtained by multiplying the characteristic matrix of each layer:

$$M_{total} = M_1 M_2 ... M_N = \begin{bmatrix} M_{11} & M_{12} \\ M_{21} & M_{22} \end{bmatrix}, \quad (5)$$

based on the characteristic matrix of the whole composite structure, the structure's $M$ matrix can be obtained as follows:

$$\begin{bmatrix} B \\ C \end{bmatrix} = M_{total} \begin{bmatrix} 1 \\ \eta_{N+1} \end{bmatrix}, \quad (6)$$

by defining the optical admittance $Y = C/B$ of the structure, the reflection coefficient $R$ of the composite structure can be obtained:

$$\begin{cases} r = \dfrac{\eta_0 - Y}{\eta_0 + Y}, \\ R = |r|^2 \end{cases} \quad (7)$$

based on the transfer matrix method, on the condition of narrow spectrum Gaussian pulse, it is easy to calculate the reflected group delay as follows:

$$\tau_r = \left[\partial\varphi_r / \partial\omega\right]_{\omega=\omega_c}, \quad (8)$$

in the following calculation, we set the initial conditions, Fermi level $E_F = 0.3$ eV, electron relaxation time $\tau = 0.9$ ps, the thickness of BDS layer $d_{BDS} = 20$ nm, and the dielectric constant and thickness of the spacer layer are consistent with that of dielectric layer B.

## 3. Results and Discussions

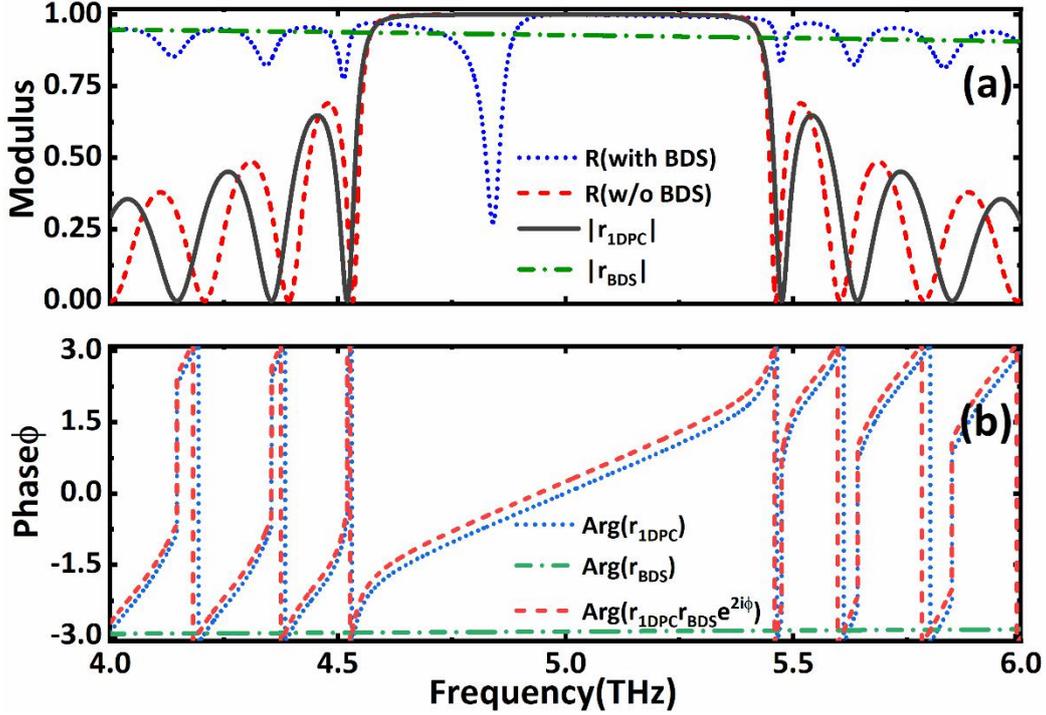

Fig.2. (a) Dependences of the reflection spectra on frequency for different structures;

(b) Dependences of the reflected phase on frequency

Firstly, we discuss the reflectivity of photonic crystal composite structure with or without BDS, 1D PC and separate BDS layer, as shown in Fig.2. It is obvious from the figure that when there is no BDS, the reflectivity of the photonic crystal is almost 1 in a certain frequency range due to the band gap, which indicates that the photonic crystal has a blocking effect on the photons in this frequency band, forming the photonic band gap. The addition of BDS makes the curve of reflectivity appear an obvious dip (nearby $f = 4.84$ THz ) in the photonic band gap. When $f = 4.84$ THz is obtained from Fig. 2, the reflection coefficient $r_{BDS}$ of electromagnetic wave incident from the spacer layer to BDS and the reflection coefficient $r_{DBR}$ of electromagnetic wave incident from the spacer layer to photonic crystal can be found to be satisfied

$r_{BDS}r_{DBR}\exp(2i\delta) \approx 1$, $Arg(r_{BDS}r_{DBR}\exp(2i\delta)) \approx 0$. Therefore, it can be considered that the BDS-1D PC composite structure excites the OTS nearby $f = 4.84$ THz [31]. In addition, at the excitation frequency of OTS, the real part of the reflection coefficient tends to zero and its imaginary part changes obviously monotonically, so the reflected phase will also monotonically change. This variation creates conditions for us to obtain large group delay at the excitation frequency of OTS.

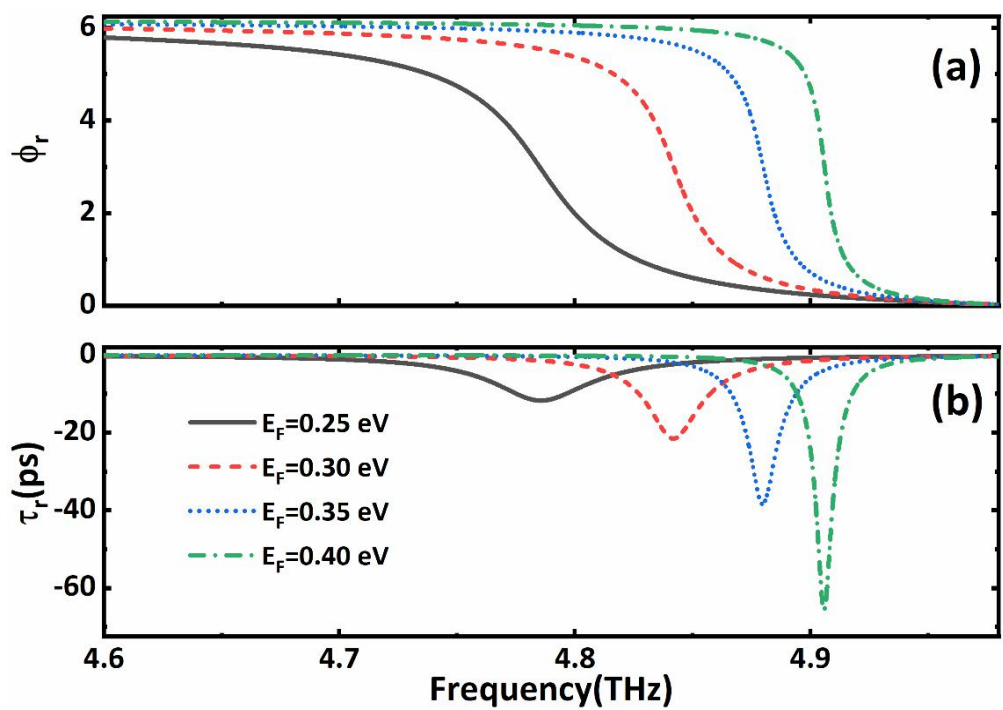

Fig.3. Dependences of the (a) reflected phase and the (b) reflected group delay on frequency of composite structures for different Fermi levels

According to formula (4), small changes of the conductivity of BDS can realize the dynamic regulation of large group delay, while the conductivity of BDS can be regulated by its Fermi level. Therefore, we need to consider the regulation ability of Fermi level on the reflected group delay. In the case of TE polarization, Fig. 3 shows the dependences between the reflected phase and reflected group delay on frequency of the composite structure for different Fermi levels. According to formulas (1) and

(3), when the Fermi level changes, the conductivity of BDS also changes, which further affects the dielectric constant of BDS, and finally obtains different group delays. Therefore, by applying external voltage to change the Fermi level of BDS, we can dynamically adjust the value of group delay. When the Fermi level increases, the point of resonant frequency shifts to the right, and the curve of the reflected phase further becomes steep, as shown in Fig. 3 (a). The change of the phase is directly reflected on the value of the group delay. As shown in Fig. 3 (b), it can be found that when the Fermi level is $0.25$ eV, the group delay of about -11.75 ps can be obtained. Continuing to increase the Fermi level can help us obtain a greater negative group delay. For example, when the Fermi level is $0.4$ eV, the group delay is about -65.7 ps. At the same time, the excitation frequency of OTS also moves towards a higher frequency with the increase of the Fermi level. These phenomena show that the reflected group delay of the composite structure is very sensitive to the change of the Fermi level, Fermi level plays an important role in determining the value of group delay. This electronic control characteristic of the reflected group delay provides a way to design tunable group delay devices with high sensitivity. For the convenience of discussion, we will discuss the fixed Fermi level of $0.4$ eV.

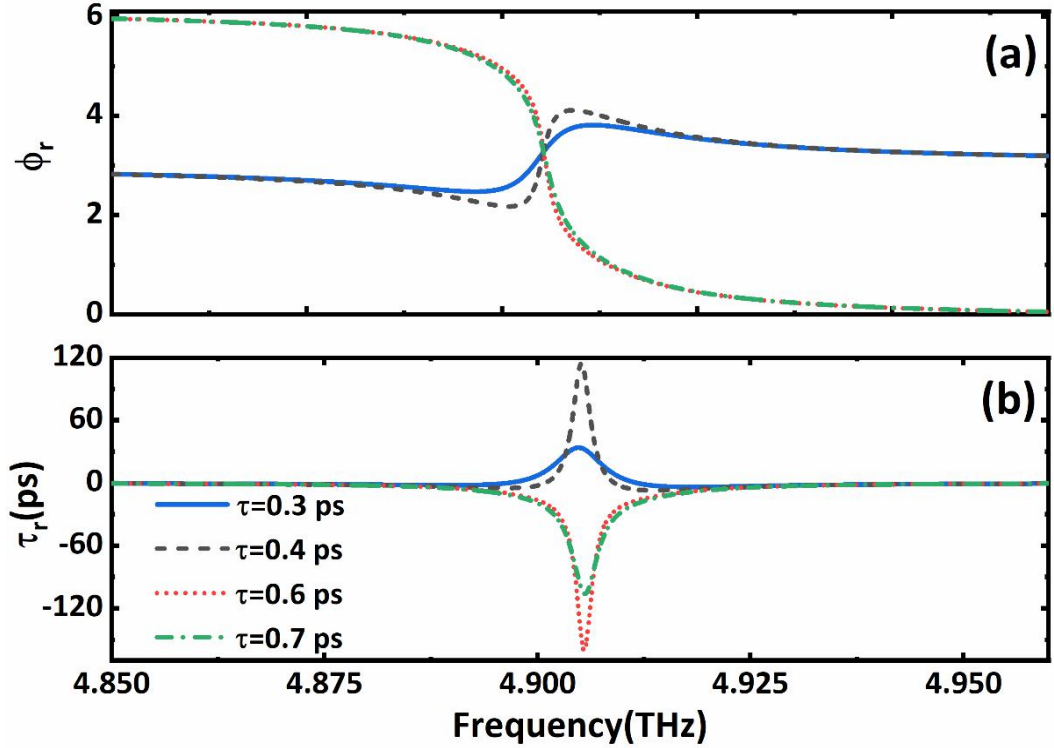

Fig.4. Dependences of the (a) reflected phase and the (b) reflected group delay on frequency for different electron relaxation time

Apart from the dynamic regulation of the group delay by Fermi level, it is known from formula (3) that the electron relaxation time of BDS also has a significant impact on the conductivity of BDS. According to the calculation by the transfer matrix method, the reflected group delay will also be very sensitive to the electron relaxation time. This characteristic provides a new way to regulate the reflection pulse group delay. Fig. 4 shows the change of the reflected group delay and the reflected phase on frequency for different electron relaxation time. Compared with the influence of Fermi level on the reflected phase and the reflected group delay, the influence of electron relaxation time on the reflected group delay is mainly illustrated in the value of the peak of group delay, and the frequency (resonant frequency) of the reflected phase and the reflected group delay will not be changed. Further, we find that the change of the electron relaxation time of BDS can change the monotonicity of the reflected phase, thus changing the positive and negative of the reflected group delay. It is worth noting that compared with the positive and negative jump of the reflected group delay of grapheme- photonic crystal structure when the electron relaxation time approaching infinity, the composite structure discussed in this paper is the jump at the smaller electron relaxation time. As can be seen from Fig. 4 (a) (b), when the ratio of Fermi level to electron relaxation time is greater than 1, a positive group delay can be obtained, and with the increase of relaxation time, the reflected phase becomes steeper near the resonant frequency and the reflected group delay increases; when the ratio of Fermi level to electron relaxation time is less than 1, the sign of group delay is reversed. At this time, the value of group delay becomes smaller and smaller with the increase of relaxation time. This regulation characteristic is of great value, it provides a way for the conversion of positive and negative reflected group delay. It should be noted that for a certain structure, it is very difficult to change the electron relaxation time of the structure. However, for the BDS-1D PC composite structure, according to the electron relaxation time expression of BDS, the relaxation time is regulated by the Fermi level, so the relaxation time can also be regulated by adjusting the external voltage. This composite structure has unique advantages over various previous structures that realize delay.

Next, we discuss the effect of parameters of BDS-1D PC composite structure on the overall reflected group delay. These conclusions will provide key references for the design of reasonable group delay devices. Fig. 5 shows a multicolor diagram of the group delay of the thickness of the BDS dielectric layer on different frequencies and the dependences of the maximum group delay value on the thickness of the BDS layer. Because the thickness of the dielectric layer directly affects the characteristic matrix of the medium and further affects the reflected group delay, the reflected group delay is very sensitive to the change of the thickness of the dielectric layer. As can be seen from Fig. 5 (a) (b), when the thickness of other dielectric layers remains unchanged, appropriately increasing the thickness of BDS layer can significantly increase the value of negative group delay, and the excitation frequency of OTS moves towards high frequency. However, when the thickness of BDS layer increases to a certain value (greater than 47 nm ), the continuous increase of $d_{BDS}$ will lead to the jump of group delay from negative to positive. If $d_{BDS}$ continues to increase when $d_{BDS} > 47 \text{ nm}$, the positive reflected group delay will be significantly reduced. When $d_{BDS}$ reaches a certain value, the value of reflected group delay and the excitation frequency of OTS will tend to be stable. Therefore, we can draw a conclusion that an appropriate BDS layer thickness is a necessary condition to obtain a large reflected group delay value, which must be considered in the design of tunable group delay devices.

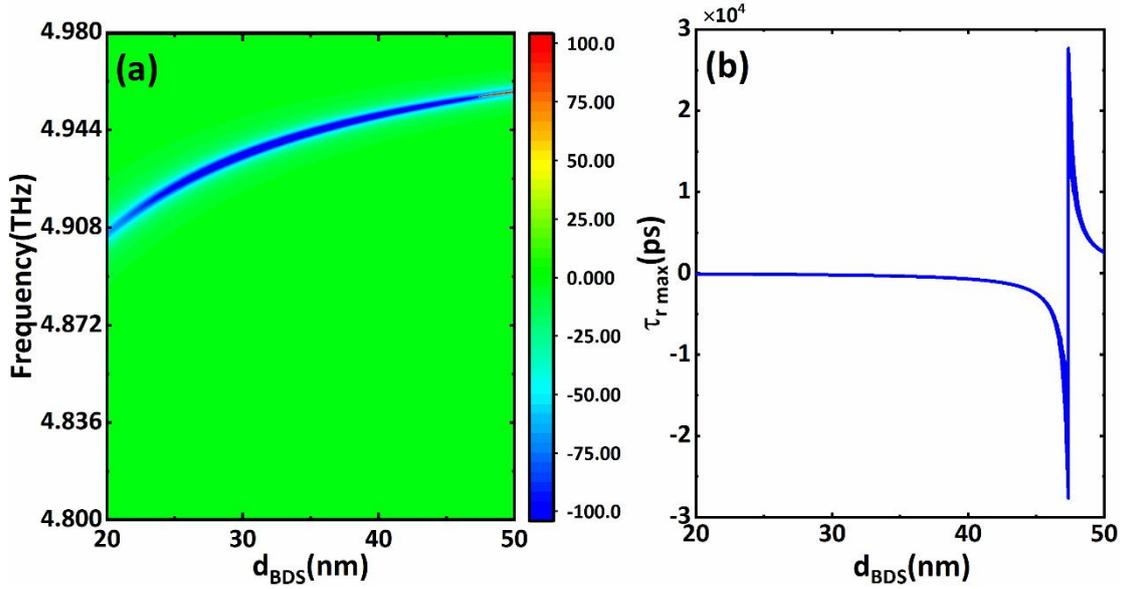

Fig.5. (a) Colorful diagram of group delay of BDS dielectric layer thickness on different frequencies; (b) Dependences of the maximum group delay value on the thickness of BDS layer

## 4. Conclusions

In conclusion, we studied the regulation of the reflected group delay of the terahertz band vertical incident pulse when the OTS is excited in the BDS-1D PC structure. The simulation results show that the sharp phase change of the OTS excited on BDS-1D PC is the main factor to enhance and tune the reflected group delay. The regulation of group delay mainly depends on the conductivity characteristics of BDS at the excitation frequency of OTS (such as the conductivity change of BDS). At the same time, we also found that the group delay can not only be regulated by the external voltage which leads to the change of the Fermi level, but can also obtain a huge group delay by changing the parameters of the composite structure. These findings provide a feasible way to control the delay characteristics of terahertz pulses. We believe that the BDS-1D PC is helpful to obtain huge tunable group delay and is of great significance in the design and preparation of controllable group delay devices with high sensitivity.

pp. 5417, 2016.